\documentclass[useAMS,usenatbib,usegraphicx]{mn2e}
\usepackage{times}
\usepackage{longtable}
\usepackage{amsmath}
\usepackage{amssymb}
\usepackage{nopageno}
\usepackage{longtable}
\usepackage{graphicx}
\usepackage{pifont}

\def\hatlas{{\it H}-ATLAS}

\def\gs{\mathrel{\raise0.35ex\hbox{$\scriptstyle >$}\kern-0.6em
\lower0.40ex\hbox{{$\scriptstyle \sim$}}}}
\def\ls{\mathrel{\raise0.35ex\hbox{$\scriptstyle <$}\kern-0.6em
\lower0.40ex\hbox{{$\scriptstyle \sim$}}}}
\def\m@th{\mathsurround=0pt }
\def\eqalign#1{\null\,\vcenter{\openup1\jot \m@th
 \ialign{\strut\hfil$\displaystyle{##}$&$\displaystyle{{}##}$\hfil
 \crcr#1\crcr}}\,}

\title[PACS data reduction]
      {H-ATLAS: PACS imaging for the Science Demonstration Phase}
\author[Ibar et al.]
       {Edo Ibar,$^{\! 1}$\ 
         R.\,J.\ Ivison,$^{\! 1,2}$
         A.\ Cava,$^{\! 3}$
         G.\ Rodighiero,$^{\! 4}$
         S.\ Buttiglione,$^{\! 5}$
         P.\ Temi,$^{\! 6}$
         D.\ Frayer,$^{\! 7}$
         \and
         J.\ Fritz,$^{\! 8}$
         L.\ Leeuw,$^{\! 6}$
         M.\ Baes,$^{\! 8}$
         E.\ Rigby,$^{\! 9}$
         A.\ Verma,$^{\! 10}$
         S.\ Serjeant,$^{\! 11}$
         T.\ M\"uller,$^{12}$
	 \and
         R.\ Auld,$^{\! 13}$ 
         A.\ Dariush,$^{\! 13}$
         L.\ Dunne,$^{\! 9}$   
         S.\ Eales,$^{\! 13}$  
         S.\ Maddox,$^{\! 9}$  
         P.\ Panuzzo,$^{14}$ 
         E.\ Pascale,$^{\! 13}$
	 \and
         M.\ Pohlen,$^{\! 13}$ 
         D.\ Smith,$^{\! 9}$ 
         G.\ de Zotti,$^{\! 5,15}$
         M.\ Vaccari,$^{\! 4}$
         R.\ Hopwood,$^{\! 11}$
         A.\ Cooray,$^{\! 16}$
         \and
         D.\ Burgarella$^{17}$
         and M.\ Jarvis$^{18}$
         \vspace*{2mm} \\
         $^1$ UK Astronomy Technology Centre, Royal Observatory, Blackford Hill,
         Edinburgh EH9 3HJ\\
         $^2$ Institute for Astronomy, University of Edinburgh, Royal
         Observatory, Edinburgh EH9 3HJ\\
         $^3$ Instituto de Astrof\'isica de Canarias and Departamento
         de Astrof\'isica de la Universidad de La Laguna, La Laguna,
         Tenerife, Espa\~na \\ 
         $^4$ University of Padova, Vicolo Osservatorio 3, I-35122,
         Padova, Italy \\ 
         $^5$ INAF-Osservatorio Astronomico di Padova, Vicolo
         Osservatorio 5, I-35122, Padova, Italy \\
         $^6$ Astrophysics Branch, NASA Ames Research Center, MS
         245-6, Moffett Field, CA 94035, USA \\ 
         $^7$ Infrared Processing and Analysis Center, California
         Institute of Technology 100-22, Pasadena, CA 91125, USA \\
         $^8$ Sterrenkundig Observatorium, Universiteit Gent,
         Krijgslaan 281 S9, B-9000 Gent, Belgium \\
         $^9$ School of Physics and Astronomy, University of Nottingham,
         Nottingham NG7 2RD\\
         $^{10}$ Oxford Astrophysics, Denys Wilkinson Building,
         University of Oxford, Keble Road, Oxford OX1 3RH\\
         $^{11}$ Dept. of Physics and Astronomy, The Open University,
         Milton Keynes MK7 6AA\\
         $^{12}$ Max-Planck-Institut f\"ur extraterrestrische Physik,
         Giessenbachstrasse, 85748 Garching, Germany \\
         $^{13}$ School of Physics and Astronomy, Cardiff University,
         Queens Buildings, The Parade, Cardiff CF24 3AA \\
         $^{14}$ CEA, Laboratoire AIM, Irfu/SAp, F-91191
         Gif-sur-Yvette, France \\
         $^{15}$ SISSA, Via Bonomea 265, I-34136 Trieste, Italy \\
         $^{16}$ Department of Physics and Astronomy, University of
         California, Irvine, CA 92697, USA\\
         $^{17}$ Laboratoire d'Astrophysique de Marseille,
         Observatoire Astronomique Marseille Provence, Aix-Marseille
         Universit\'e, CNRS, France\\
         $^{18}$ Centre for Astrophysics Research, STRI, University of
         Hertfordshire, Hatfield AL10 9AB\\ 
       }
\date{Accepted 2010 August 27.  Received 2010 August 26; in original form 2010 June 8}
\pagerange{000--000}

\begin{document}

\maketitle

\begin{abstract} We describe the reduction of data taken with the PACS
  instrument on board the {\it Herschel Space Observatory} in the
  Science Demonstration Phase of the {\it Herschel}-ATLAS (\hatlas)
  survey, specifically data obtained for a 4$\times$4-deg$^2$ region
  using {\it Herschel}'s fast-scan (60\,arcsec\,s$^{-1}$) parallel
  mode. We describe in detail a pipeline for data reduction using
  customised procedures within HIPE from data retrieval to the
  production of science-quality images. We found that the standard
  procedure for removing Cosmic-Ray glitches also removed parts of
  bright sources and so implemented an effective two-stage process to
  minimise these problems.  The pronounced $1/f$ noise is removed from
  the timelines using 3.4- and 2.5-arcmin boxcar high-pass filters at
  100 and 160\,$\mu$m. Empirical measurements of the point-spread
  function (PSF) are used to determine the encircled energy fraction
  as a function of aperture size. For the 100- and 160-$\mu$m bands,
  the effective PSFs are $\sim$9 and $\sim$13\,arcsec ({\sc fwhm}),
  and the 90-per-cent encircled energy radii are 13 and
  18\,arcsec. Astrometric accuracy is good to $\ls$2\,arcsec. The
  noise in the final maps is correlated between neighbouring pixels
  and rather higher than advertised prior to launch. For a pair of
  cross-scans, the 5-$\sigma$ point-source sensitivities are
  125--165\,mJy for 9--13-arcsec-radius apertures at 100\,$\mu$m and
  150--240\,mJy for 13--18-arcsec-radius apertures at 160\,$\mu$m.
\end{abstract}


\section{Introduction}

The 3.5-m {\it Herschel Space Observatory}\footnote{{\it Herschel} is
  an ESA space observatory with science instruments provided by
  European-led Principal Investigator consortia and with important
  participation from NASA.}  \citep{pilbratt10} is the first space
telescope to cover the entire far-infrared waveband (from 55 to
670\,$\mu$m) and looks likely to become one of the greatest
astronomical achievements of this decade.

The {\it Herschel} Astrophysical Terahertz Large Area Survey
\citep[\hatlas\ --][]{eales10} is the largest {\it Herschel} Open Time
Key Project (600\,hr), covering 550\,deg$^2$ of sky in regions
selected on the basis of existing multi-wavelength coverage (e.g.\ the
Galaxy Evolution Explorer -- GALEX, the Galaxy and Mass Assembly
spectroscopic survey -- GAMA, the 2dF Galaxy Redshift Survey --
2DFGRS, the Sloan Digital Sky Survey -- SDSS, and the Dark Energy
Survey -- DES). \hatlas\ will detect hundreds of thousands of galaxies
(\citealt{clements10}), and provide an extensive census of
dust-obscured activity in the local ($z<0.3$) Universe
(\citealt{amblard10}). The areal coverage of \hatlas\ also makes it
well suited to the identification of {\it Planck} sources
(e.g.\ \citealt{gonzalez-nuevo10}), lensed galaxies at high redshift
(e.g.\ \citealt{negrello07, swinbank10}) and local dust clouds at high
Galactic latitudes; there is also enormous potential for serendipitous
discovery.

\hatlas\ is exploiting the fast-scan (60\,arcsec\,s$^{-1}$) parallel
mode of {\it Herschel}, using two of the on-board instruments to
provide an efficient way to map large areas of sky in five wavebands
simultaneously. We are using the Photodetector Array Camera and
Spectrometer \citep[PACS --][]{poglitsch10} to observe at 100 and
160\,$\mu$m (its `green' and `red' channels) whilst the Spectral and
Photometric Imaging Receiver \citep[SPIRE --][]{griffin10} is taking
data at 250, 350 and 500$\,\mu$m.

This paper is one of a series of four technical papers describing our
approach to the PACS (this paper) and SPIRE (Pascale et al., in
preparation) data products, to source extraction (Rigby et al., in
preparation) and to cross-identification (Smith et al., in
preparation) for the Science Demonstration Phase (SDP) of the
\hatlas\ survey. These data are public and available at {\tt
  http://www.h-atlas.org/}. Here, we describe the pipeline used to
reduce data obtained with PACS, and the quality of data products, to
give an idea of its scientific potential. In \S\ref{Sinstrument}, we
provide a brief description of the instrument; in \S\ref{Sthedata}, we
present the \hatlas\ SDP observations; in \S\ref{Spipeline}, we
describe the customised procedures we have developed within the {\it
  Herschel} Interactive Processing Environment (HIPE\footnote{HIPE is
  a joint development by the {\it Herschel} Science Ground Segment
  Consortium, consisting of ESA, the NASA {\it Herschel} Science
  Center, and the HIFI, PACS and SPIRE consortia. HIPE is a graphical
  application which includes Jython scripting, data analysis,
  plotting, communication with the {\it Herschel} Science Archive and
  much more. Throughout the paper we refer to version 3.0.859, which
  is the build we used to reduce the \hatlas\ PACS SDP data. Note that
  HIPE is under continuous development.} -- \citealt{ott10}) to reduce
data from PACS. In \S\ref{Sanalyses} we describe tests of the
resulting images and we state some concluding remarks in
\S\ref{Sconclusions}.

\section{PACS instrument}
\label{Sinstrument}

PACS is a multi-colour camera and low- and medium-resolution
spectrometer covering the 55--210-$\mu$m wavelength range (see
Fig.~\ref{filters}). It comprises four large-format detector arrays:
two filled silicon bolometer arrays optimised for imaging in
high-photon-background conditions and two Ge:Ga photo-conductor arrays
for spectroscopy. Here we concentrate on the bolometer detectors used
by the \hatlas\ survey -- a more complete description of the
instrument and its modes can be found at \citet{poglitsch10}. A
dichroic beam splitter enables photometry in two bands simultaneously
-- 70 or 100\,$\mu$m (`blue' or `green', selected by a filter wheel)
and 160\,$\mu$m (`red') -- over the same 1.75$\times$3.5-arcmin$^2$
field of view. The bolometer arrays comprise 64$\times$32 and
32$\times$16 pixels, with 3.2 and 6.4\,arcsec pixel$^{-1}$ on-sky,
respectively, providing close-to Nyquist beam sampling for the
blue/green and red filters.  The arrays each comprise sub-arrays of
16$\times$16 pixels, tiled together to form the focal plane (see
\citealt{billot09} and references therein).

Working in `scan mode', PACS modulates the sky signal by making use of
the motion of the spacecraft (10, 20 or 60\,arcsec\,s$^{-1}$), with no
chopping. The sky signal is stored in units of mV by the
analogue-to-digital (ADU) converter, depending on the user-defined
V/ADU gain value (high or low; see \citealt{poglitsch10}).  The signal
from each bolometer pixel is sampled at a rate of 40\,Hz, although due
to satellite telemetry limitations, in scan mode the signal is
averaged into packages of four consecutive frames -- i.e.\ resulting
in an effective rate of 10\,Hz. Data is also bit-rounded by the
Signal-Processing Unit (SPU) (\citealt{ottensamer08}) which results in
a stronger quantisation of the signal than would be expected by the
ADU converter. When using the SPIRE/PACS `fast-scan parallel mode', as
employed by \hatlas, in particular for the blue/green filters, data
suffer additional in-flight averaging (eight frames), resulting in an
effective read-out frequency of 5\,Hz.

Due to the limited signal bandwidth of the detection chain, the
on-board data compression produces significant degradation of the
observed point spread function (PSF; Instrument Control Centre -- ICC
report\footnote{PICC-NHSC-TR-011, June 2008, version 0.3. Report by
  N.\ Billot et al.\ \label{fnbillot}}). Simulated parallel mode data
based on PACS prime fast-scan observations show that point-source peak
fluxes are reduced by $\sim$50 and 70 per cent, at 100 and 160\,$\mu$m
respectively, in comparison to nominal (20\,arcsec\,s$^{-1}$) scan
observations\footnote{\label{fnlutz}PICC-ME-TN-033, November 10, 2009,
  version 0.3 report by D.\ Lutz} (\citealt{poglitsch10}).

Prior to assessment of the in-flight performance, the predicted
5-$\sigma$ point-source \hatlas\ sensitivities based on the {\it
  Herschel}-Spot (HSpot\footnote{\tt
  herschel.esac.esa.int/Tools.shtml}) observation planning tool were
67 and 94\,mJy for observations with one pair of cross-scans, at 100
and 160\,$\mu$m, respectively (\citealt{eales10}). The performance is
mostly dependent on the optical efficiency, the thermal and telescope
background, the effects of Cosmic-Ray glitches -- in particular
high-energy protons -- and the photon noise from the detector and
multiplexer electronics which was found to introduce a $1/f$ excess
below 1\,Hz prior to launch (see \S\,\ref{Sprojection}).

The PACS focal plane is offset with respect to SPIRE by a fixed
separation of $\sim$21\,arcmin along the spacecraft $z$-axis, implying
different instantaneous PACS and SPIRE coverages. The coverage
obtained by SPIRE in the SDP area is presented in
\S\,\ref{Sprojection} and clearly show that parallel mode is only
efficient for large surveyed areas. For this mode, the angle between
the spacecraft $z$-axis and the scan direction is either +42.4 or
$-$42.4\,deg in order to obtain an uniform coverage for SPIRE (see
PACS Observer's Manual\footnote{\tt
  herschel.esac.esa.int/Docs/PACS/pdf/pacs\_om.pdf})

\section{PACS \hatlas\ SDP data}
\label{Sthedata}

On 2009 November 22 (Observing Day -- OD\,192) {\it Herschel} observed
one of the equatorial fields of the \hatlas\ survey (see Table~1 of
\citealt{eales10}). Observations covered an area of approximately
4$\times$4\,deg$^2$ (a quarter of the GAMA-A -- also called the
GAMA-9h -- field). These constitute the \hatlas\ SDP observations (see
Table~\ref{obs_table}).

\begin{figure}
   \centering
   \begin{tabular}{cc}
    \includegraphics[width=8.2cm]{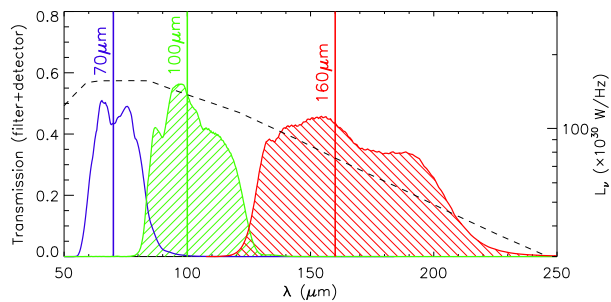}
   \end{tabular}
   \caption{Transmission filter/detector profiles for the PACS 55--85-
     (blue), 85--125- (green) and 125--210-$\mu$m (red) passbands. The
     filled profiles show the filters used for the
     \hatlas\ observations (green and red). The dashed line shows the
     Spectral Energy Distribution (SED) of M\,82 with units on the
     right-hand axis.}
   \label{filters}
\end{figure}

Of the two available combinations of photometric bands for PACS, we
opted to observe at 100 and 160\,$\mu$m (see Fig.~\ref{filters})
because these are best suited to our science goals. Approximately
44\,Gbytes of data were retrieved using HIPE via the {\it Herschel}
Science Archive (HSA) interface.

\begin{table*}
\centering
\begin{tabular}{|ccccccccc|}
\hline
\hline
Proposal &
Target &
Scan &
R.A. (J2000) &
Dec. (J2000) &
Observing date &
UT start time &
Duration  &
OBSID\\
 &
 &
 &
(h:m:s) &
($^\circ$:'':') &
(y-m-d) &
(h:m:s) &
(h) &
 \\
\hline
SDP\_seales01\_6 &
ATLAS\_SDn &
Nominal &
09:05:30.0 &
+0:30:00.0 &
2009-11-22 &
00:18:02.392 &
8.11261 & %28461 &
1342187170 \\
SDP\_seales01\_6 &
ATLAS\_SDn &
Orthogonal &
09:05:30.0 &
+0:30:00.0 &
2009-11-22 &
08:24:55.277 &
8.05333 & %28461 &
1342187171 \\
\hline
\hline
\end{tabular}
\caption{The fast-scan parallel mode SDP observations
  (4$\times$4\,deg$^2$) for the \hatlas\ survey. \label{obs_table} }
\end{table*}

One pair of cross-scans were taken, covering the entire
4$\times$4\,deg$^2$ field. The two resulting datasets (Observation ID
-- OBSID 1342187170 and 1342187171) comprise 89 and 97 parallel
scan-legs, respectively, each $\sim$4\,deg in length, separated from
each other by $\sim$2.6\,arcmin. After completing a scan-leg, the
telescope turns around and performs the next parallel scan in the
opposite direction.  Every ten scan-legs, approximately, calibration
observations are made to track any drifts in detector response: the
telescope chops in a stationary position at the edge of the scan-leg
for approximately half a minute (\citealt{krause06}). Given the
in-flight stability of the PACS bolometers, the {\it Herschel} Science
Centre (HSC) has decided to restrict calibration blocks to one per
OBSID for the rest of \hatlas\ observations.

As well as ensuring good coverage of the field, the acquisition of two
independent cross-scan measurements allows us to identify and remove
drifts in the data timelines\footnote{{\tt
    herschel.esac.esa.int\\ /Docs/PMODE/html/parallel\_om.html}} and
to use maximum-likelihood imaging algorithms
(e.g.\ \citealt{cantalupo10}, \citealt{patanchon08}) which can help to
mitigate the strong $1/f$ noise present in the data (see
\S\,\ref{SSalternative}).

\section{The pipeline}
\label{Spipeline}

Our data analysis faced a significant computational challenge. To
process the full pipeline and given by the particularly large data
set, one must set up HIPE to increase the available random-access
memory (RAM) limits to 60\,Gbytes (or more). Within HIPE, the pipeline
is written in Jython (a Python implementation written in Java) and
consists of a series of tasks developed by the PACS ICC in
collaboration with the HSC. HIPE allows the user to specify
data-reduction steps, from data retrieval to the final imaging
processes.

\begin{figure}
  \centering
  \includegraphics[width=8.2cm, height=13cm, clip=true, trim= 2.5cm 1cm 2.5cm 1.5cm]{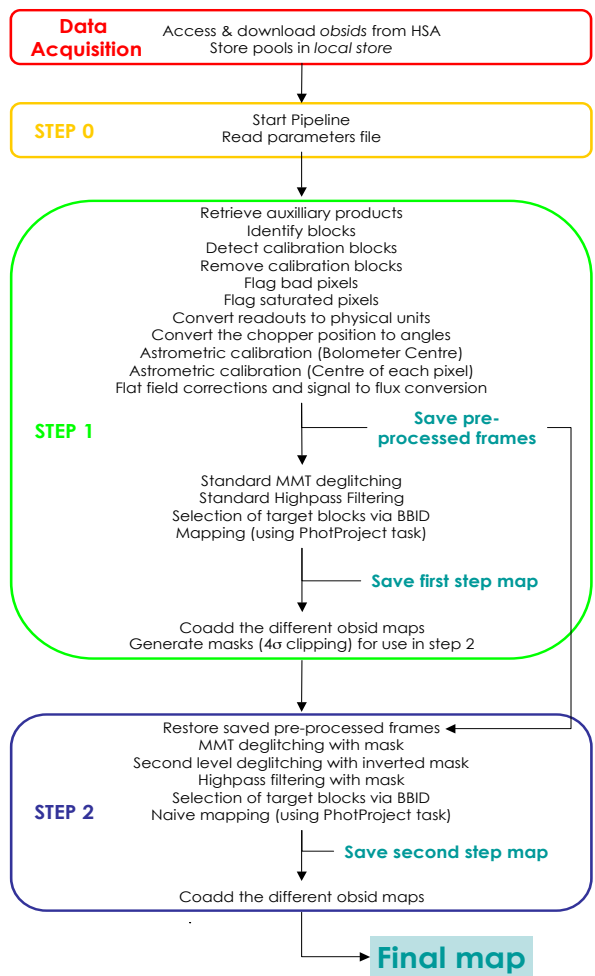}
   \caption{A flow-chart of the pipeline used for the PACS data
     reduction.}
   \label{Fpipeline}
\end{figure}

Thanks to the development of HIPE, the data processing is relatively
straightforward. Nevertheless, it needs to be fine-tuned to deal
effectively with cosmic-ray (CR) removal (deglitching -- see
\S\,\ref{SSdeglitching}), and to perform the imaging necessary to
tackle $1/f$ noise. A schematic view of the full pipeline is shown in
Fig.~\ref{Fpipeline}.

\subsection{Retrieve and organise the data}

The raw (so-called `level-0') data were retrieved using HIPE task {\tt
  getObservation} and saved into the local pool. These data contain
all the information necessary for their reduction -- science data as
well as all the necessary housekeeping/auxiliary/calibration data.
 
The signal detected by the bolometers as a function of time (also
named timeline frames) are extracted from the raw data and analysed
carefully throughout the pipeline. We next extract the telemetry of
the telescope contained in the pointing product, tables of the most
up-to-date calibration values based on in-flight performance tests
(extracted from the HIPE built using {\tt getCalTree}) and the
housekeeping data which tracks temperature and general
instrument/telescope status.

An organisational task ({\tt findBlocks}) is used to identify all the
different types of data within the timeline frames. In particular, we
remove the calibration blocks from the science frames using {\tt
  detectCalibrationBlock} and {\tt removeCalBlocks}, neglecting any
temporal variation of the detector responsivity.

\subsection{Flagging and calibration}

This data-processing stage is described in the PACS Data Reduction
Manual\footnote{\tt herschel.esac.esa.int/Data\_Processing.shtml}
and it is almost user-independent.

First, those bolometers identified as corrupt in ground-based tests
are removed using {\tt photFlagBadPixels}, while saturated pixels
are masked with {\tt photFlagSaturation}. These two tasks mask
approximately 2--3\,per cent of the bolometers and result in no
significant loss in areal coverage.

The raw signal measured by each bolometer ($S_{\rm ADU}$) is
quantified in steps of $2\times10^{5}$\,V as determined by the gain
and the bit-rounding applied to the fast-scan parallel-mode data. We
use {\tt photConvDigit2Volts} to transform ADU signals into
Volts. Possible cross-talk in the multiplexed read-out electronics has
not been taken into account. We set the chopper angle with {\tt
  convertChopper2Angle} (set to zero while scanning) in order to
obtain the instantaneous PACS line of sight with respect to the
spacecraft line of sight, and to facilitate the coordinate
determination for a reference pixel using the task {\tt
  photAddInstantPointing}. Then, using {\tt photAssignRaDec}, we
define the astrometric calibration for every pixel in the PACS
bolometer arrays.

Having flagged the timelines, converted the units to Volts and
calibrated astrometrically, we apply a flat-field correction using
{\tt photRespFlatfieldCorrection}. This task is used to convert the
signal into flux density units, making use all the available
calibration products from the {\it Herschel} campaign. In our data
processing, we have used the default calibration tree from the HIPE
v3.0.859 build. In particular we use the version-3 {\tt responsivity}
file that is known to be slightly biased. See later in
\S\ref{SSsensitivity}, where we describe the correction factors
required to obtain good flux calibration.

At this stage -- for the sake of efficiency -- we save the
pre-processed frames, before re-using them in the upcoming deglitching
and high-pass filtering tasks.

\subsection{Deglitching}
\label{SSdeglitching}

As already noted, the process of identifying and removing glitches
needs to be fine-tuned during the data processing. In this section we
describe the routines developed to take into account hits by CRs on
the detectors.

\subsubsection{Classes of observed glitches}

By inspecting the timelines, it is possible to identify two main types
of glitches.

\begin{figure}
   \centering
   \begin{tabular}{cc}
    \includegraphics[width=8.1cm]{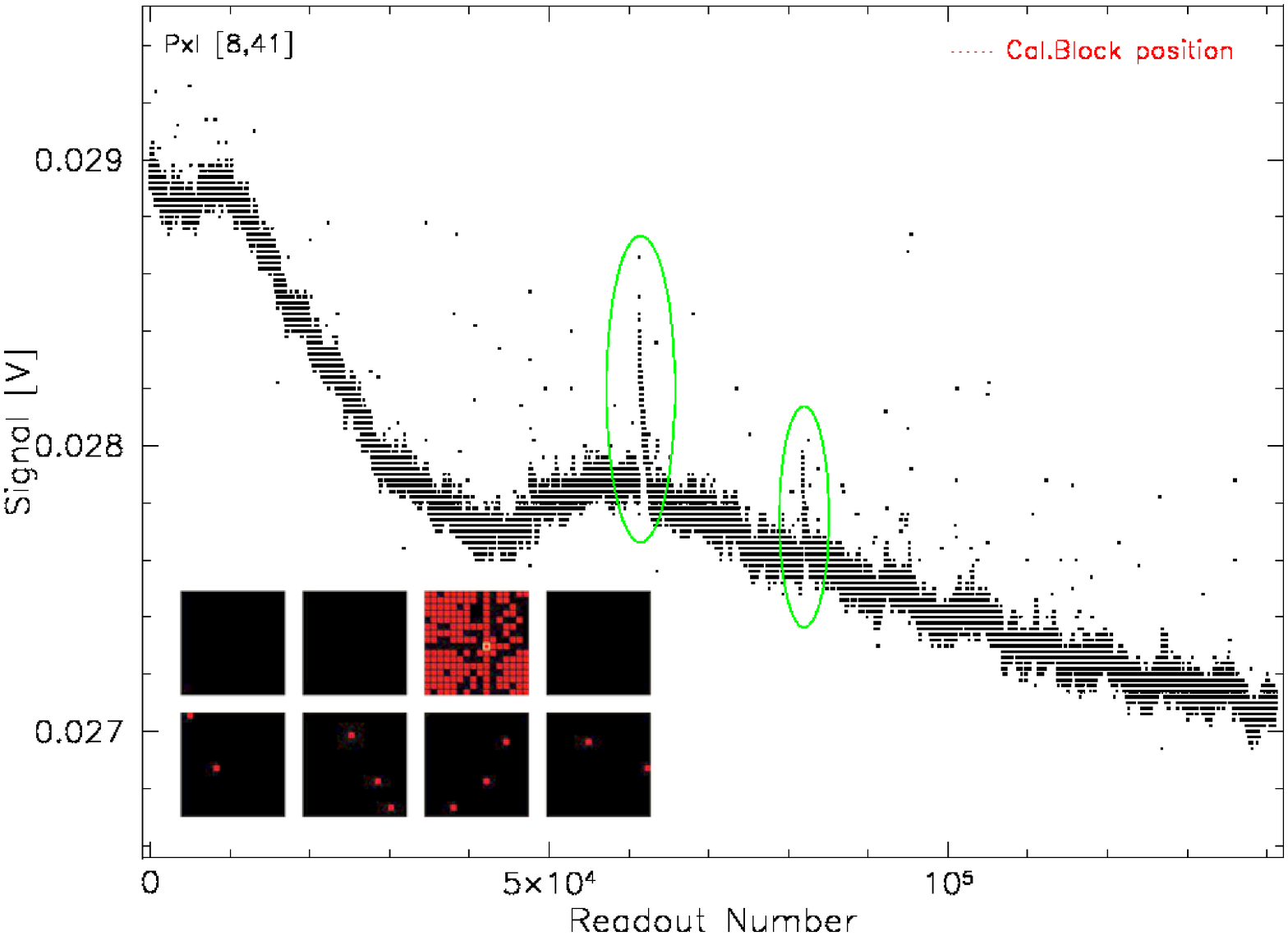} \\
    \includegraphics[width=8.1cm]{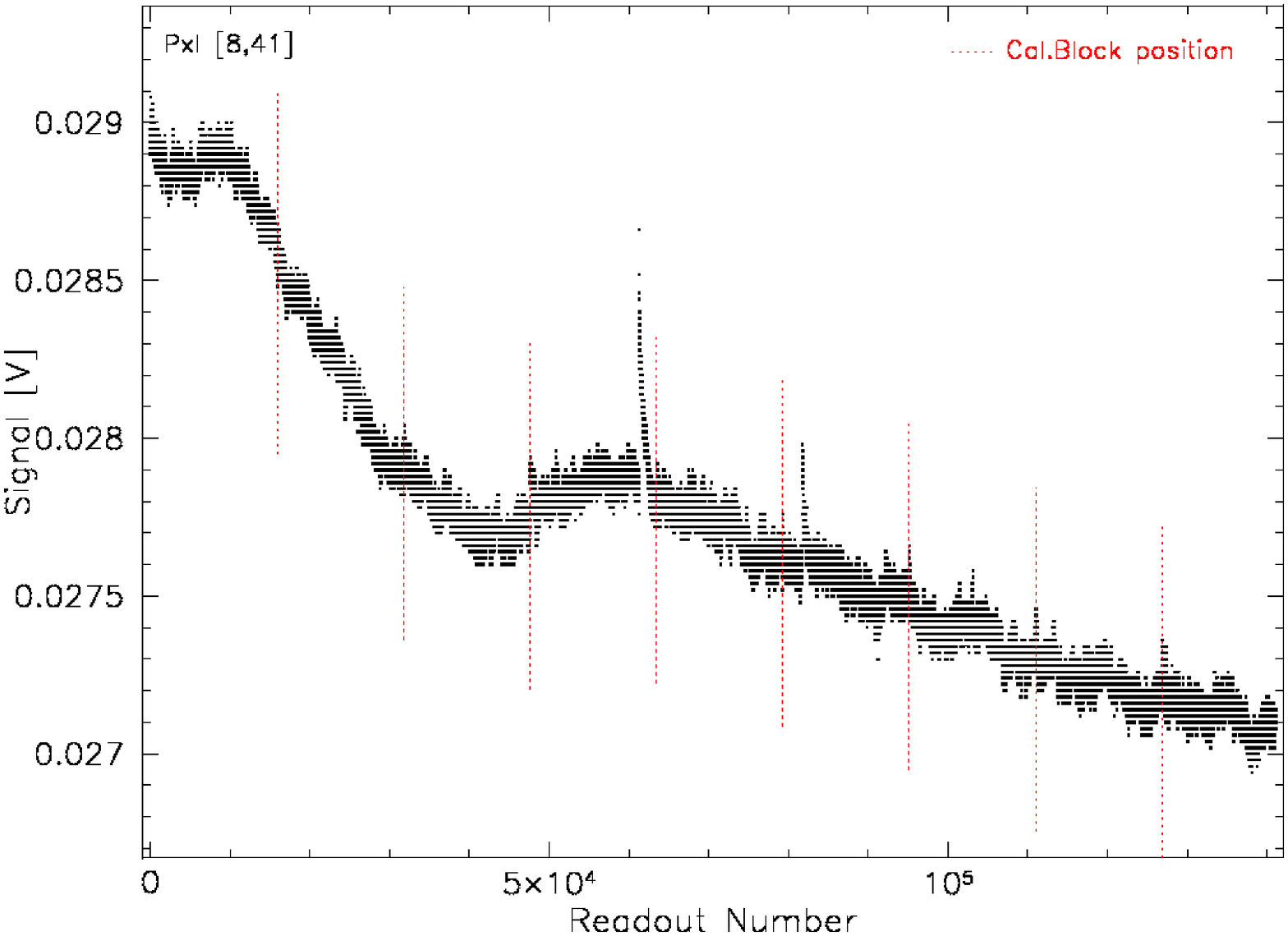} 
   \end{tabular}
   \caption{Example of a timeline for a green bolometer (8,41) in the
     \hatlas\ SDP data, before (top panel) and after (bottom panel)
     glitch removal. Green ellipses show two multi-pixel/multi-frame
     glitches. Red dotted lines show the timeline locations for
     calibration blocks (not associated to the glitches). The small
     sketch on the top figure shows the PACS bolometer pixels being
     simultaneously flagged by MMT (in red) due to a
     multi-pixel/multi-frame glitch.}
   \label{Ftimelines}
\end{figure}

First, single-pixel/single-frame glitches: these are similar to those
found in optical images, in the sense they appear in just one frame
(i.e.\ one read-out in the timeline) and they affect only one
bolometer of the detector array. We show a green bolometer's timeline
displaying such glitches in Fig.~\ref{Ftimelines} (top
panel). Glitches of this type are seen as single points, lying clearly
above (or below) the average timeline values.

In order to mask these glitches we make use of the Multi-resolution
Median Transform \citep[MMT --][]{starck98} approach (task provided
within HIPE -- {\tt photMMTDeglitching}) which performs an analysis of the
signal along each bolometer's timeline to identify outliers produced
by CR impacts. Fig.~\ref{Ftimelines} shows a typical example for the
effect that {\tt photMMTDeglitching} has on the timelines (top, original
data; bottom, data after deglitching). For our \hatlas\ observations,
we used the following parameters for this task, {\tt scales=3} and
{\tt sigma=5}, which were found to remove single-pixel/single-frame
glitches efficiently.

Secondly, multi-pixel/multi-frame glitches: these are easily
recognisable in a timeline since they are characterised by a sudden
increase (decrease) in signal, which reaches a level well above
(below) the average within one or two frames, followed by an
exponential decrease (increase) that varies in character from event to
event (see green ellipses in Fig.\ref{Ftimelines}).

The inset in the top panel of Fig.~\ref{Ftimelines} shows (in red) how
a full bolometer sub-array is affected by a multi-pixel/multi-frame
glitch (green ellipses) and masked by MMT. Such events are likely due
to a very energetic particle hitting the electronics, causing a sudden
signal increase (or decrease) in an entire sub-array for $\sim$500
frames, thus potentially affecting regions as large as $\sim$2\,deg in
the projected images. Such glitches are not fully removed by the MMT
task but are partially removed by an aggressive high-pass filter
(\S\ref{naive_proj}).

\subsubsection{Issues relating to the MMT task -- masking}
\label{SSSissues}

\begin{figure}
   \centering
   \begin{tabular}{ll}
    \includegraphics[width=4.05cm]{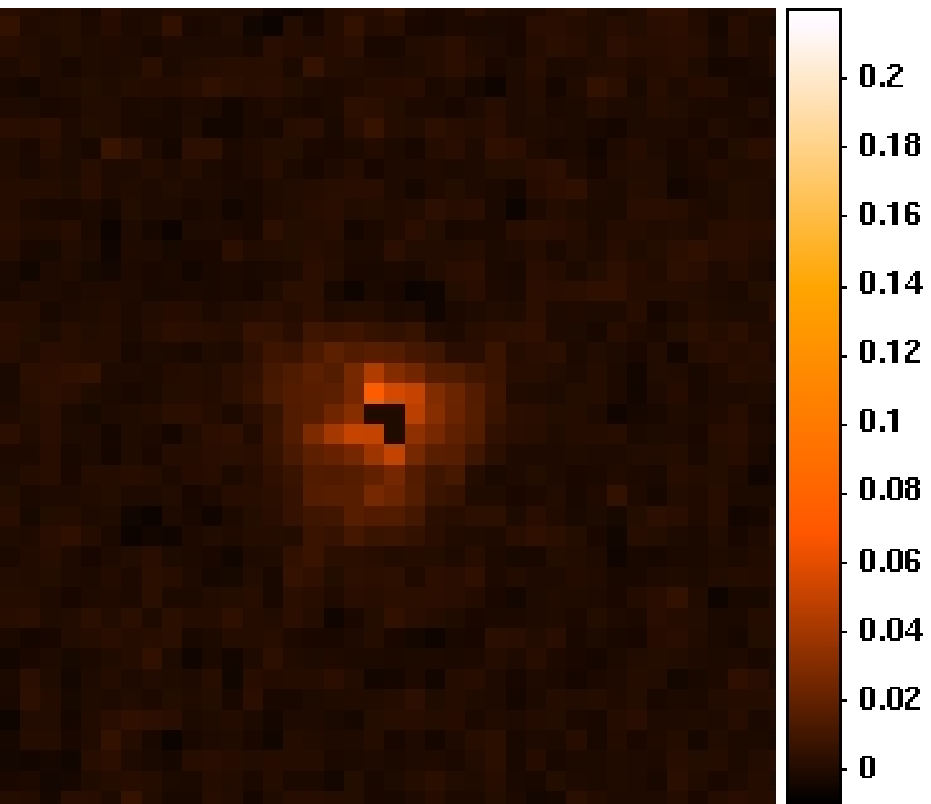} 
    \includegraphics[width=4.05cm]{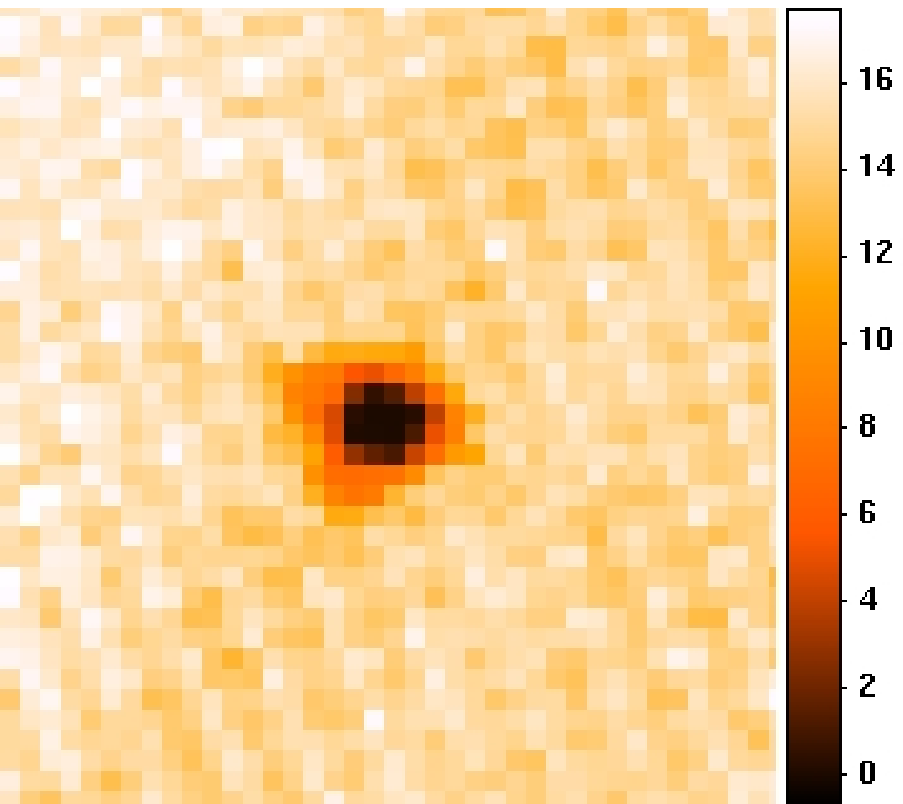}\\
    \includegraphics[width=4.05cm]{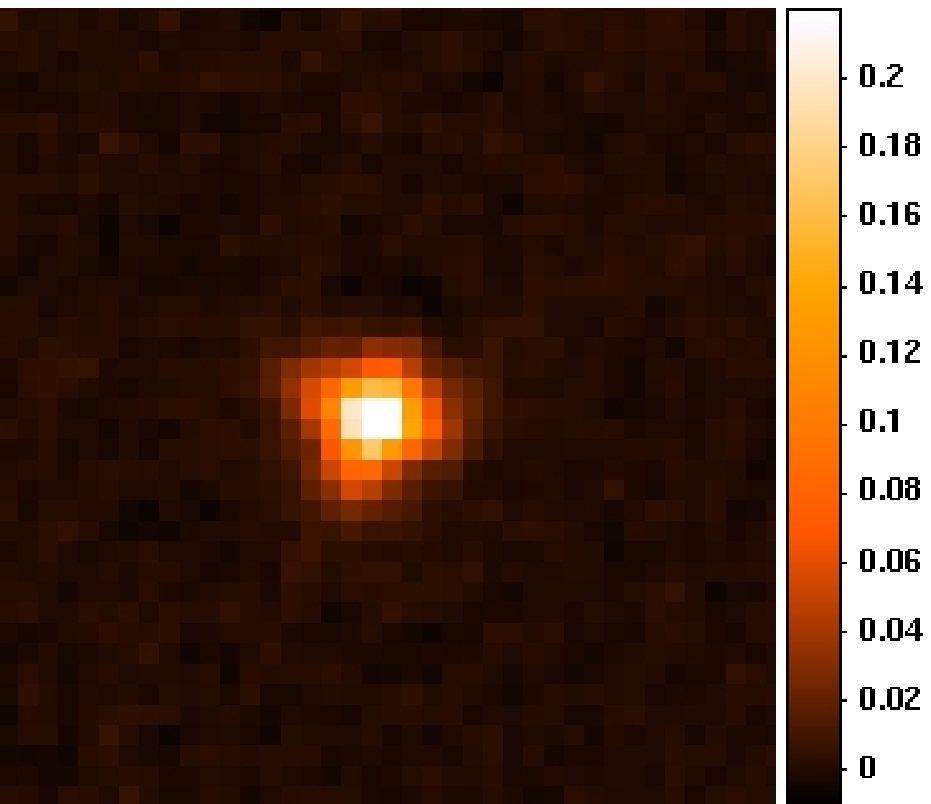} 
    \includegraphics[width=4.05cm]{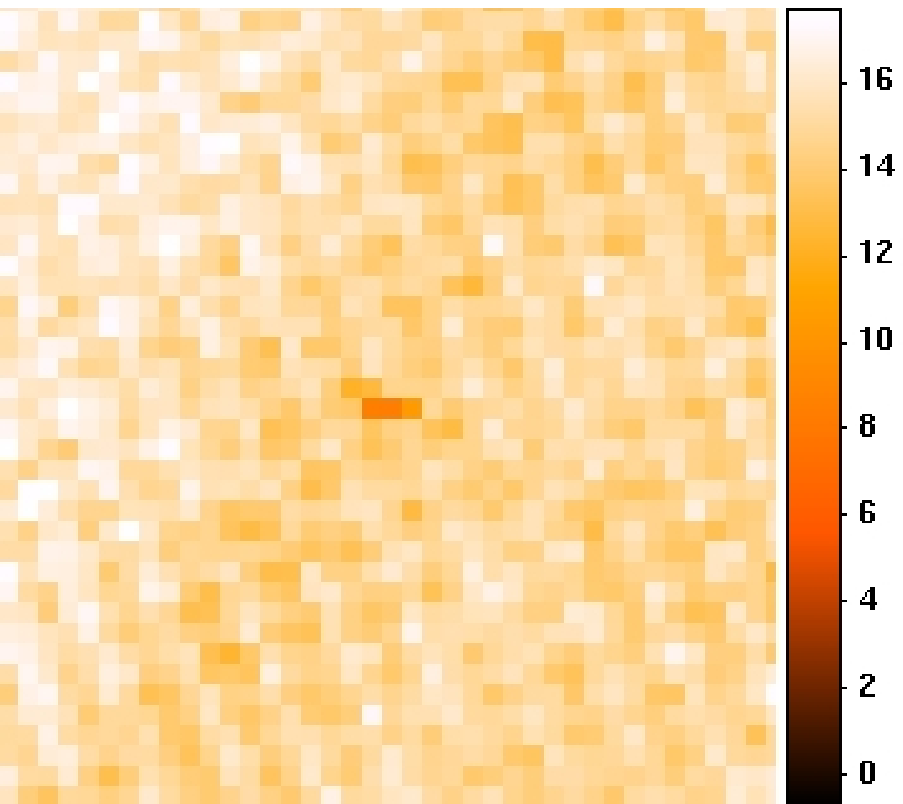} 
   \end{tabular}
   \caption{Images illustrating the importance of masking on 2nd-level
     deglitching. A bright 100-$\mu$m source image is shown to the
     left and its coverage maps (after deglitching process) to the
     right. The top panels show the removal produced by {\tt
       photMMTDeglitching}, while the bottom ones show the same source
     but using the {\tt IIndLevelDeglitchTask} approach. Image and
     coverage maps are in units of Jy/pixel (2.5\,arcsec\,pix$^{-1}$)
     and frames per pixel scales, respectively.}
   \label{fig:sara}
\end{figure}

As already noted, the use of the MMT task for deglitching is an
efficient way to remove single-pixel/single-frame glitches caused by
CRs. However, an important drawback of this algorithm is that it can
affect subsequent measurements of bright sources, i.e.\ sources whose
flux densities are significantly higher than the mean background
level. In Fig.~\ref{fig:sara} we illustrate this issue using one of
the brightest sources in the \hatlas\ SDP field. The innermost
brightest map pixels are masked and therefore they do not contribute
to the observed flux density of the source. We proved that it is
impossible to tune the {\tt photMMTDeglitching} parameters to remove
glitches effectively whilst maintaining the signal from bright
sources. Indeed, the timeline behaviour for a point source observed in
fast-scan parallel mode is almost indistinguishable from a glitch
because the effective sampling rate of 5 and 10\,Hz corresponds to
scales of 12 and 6\,arcsec when moving at 60\,arcsec\,s$^{-1}$ for the
green and red data, respectively. These scales are similar to the
observed PSFs, therefore bright point sources resemble
single-pixel/single-frame glitches -- especially in the green data.

To avoid removing flux from real sources we employ a 2nd-level stage
in the deglitching process. First, we project the timelines to
generate an image (see \S\,\ref{naive_proj}) to identify all the map
pixels where the signal is higher than 4\,$\sigma$ (the pixel
r.m.s.\ is determined from a 0.2\,deg$^2$ region devoid of bright
sources). Next, we use {\tt photReadMaskFromImage} to create a mask
for all those frames contributing at these masked positions. This mask
allows us to run {\tt photMMTDeglitching} again, this time avoiding
all those frames that contribute to the masked pixels, thus avoiding
real sources. In this way, MMT removes the vast majority of the
low-energy CR hits from the data.

Glitches powerful enough to create $>$4\,$\sigma$ events in the
aforementioned image can be identified easily as outliers in the
signals contributing to each individual pixel. We built an index map
cube with {\tt photProject} ({\tt slimindex = False}) to identify the
frames contributing to each map pixel (there are $\sim$15 and $\sim$30
frames per pixel for the 100- and 160-$\mu$m images, respectively),
and make use of {\tt IIndLevelDeglitchTask} to remove clear outliers
from these contributions (we use {\tt deglitchvector =
  'framessignal'}). In {\tt IIndLevelDeglitchTask}, the glitch
detection threshold is set to 5\,$\sigma$ using the {\tt Sigclip}
task.

To determine whether deglitching is removing flux from bright sources
we employ a simple sanity check: we examine the coverage map at the
those positions. In Fig.~\ref{fig:sara} ({\it right}), we show the
resultant coverage maps after a simple run of {\tt photMMTDeglitching}
(top) and after applying 2nd-level deglitching aided by masks
(bottom). The hole seen in the coverage represents data that have been
removed by mistake during the deglitching phase. Using 2nd-level
deglitching, this hole is significantly reduced.

We note this deglitching approach is still under development. We find
that a small amount of data are still flagged erroneously (or leaving
few remaining glitches) due to the high scatter produced by steep flux
gradients when comparing the contributions to the very brightest map
pixels.

As we have shown, we find that an effective way to treat glitches is
with a combination of the two aforementioned tasks: {\tt
  photMMTDeglitching} on the background, followed by {\tt
  IIndLevelDeglitchTask} on the possible sources, aided by source
masks in both cases -- see Fig.~\ref{Fpipeline} as our recommended
guide.

\subsection{Imaging}
\label{Sprojection}

\begin{figure}
   \centering
   \begin{tabular}{ll}
    \includegraphics[width=8.1cm]{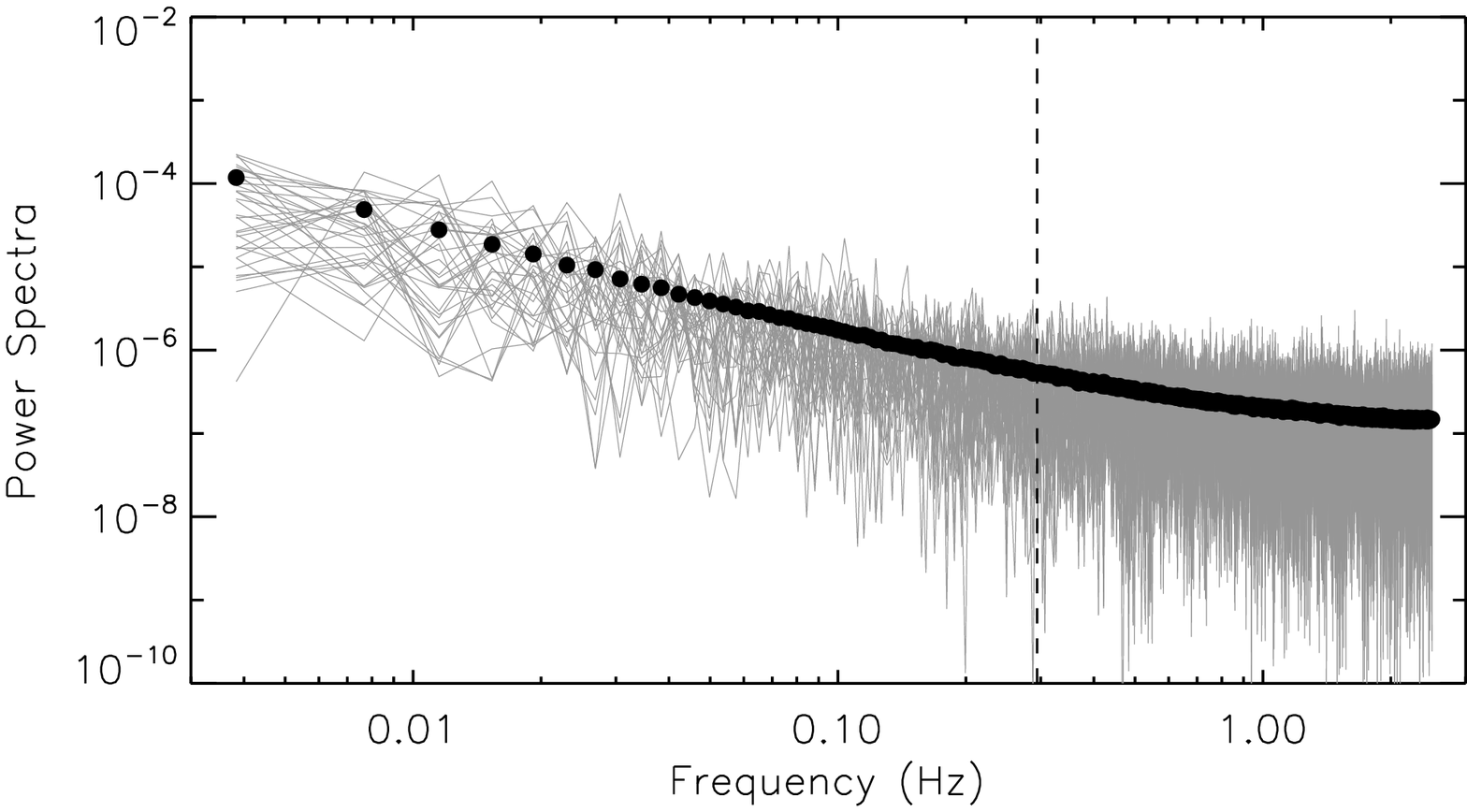} \\
    \includegraphics[width=8.1cm]{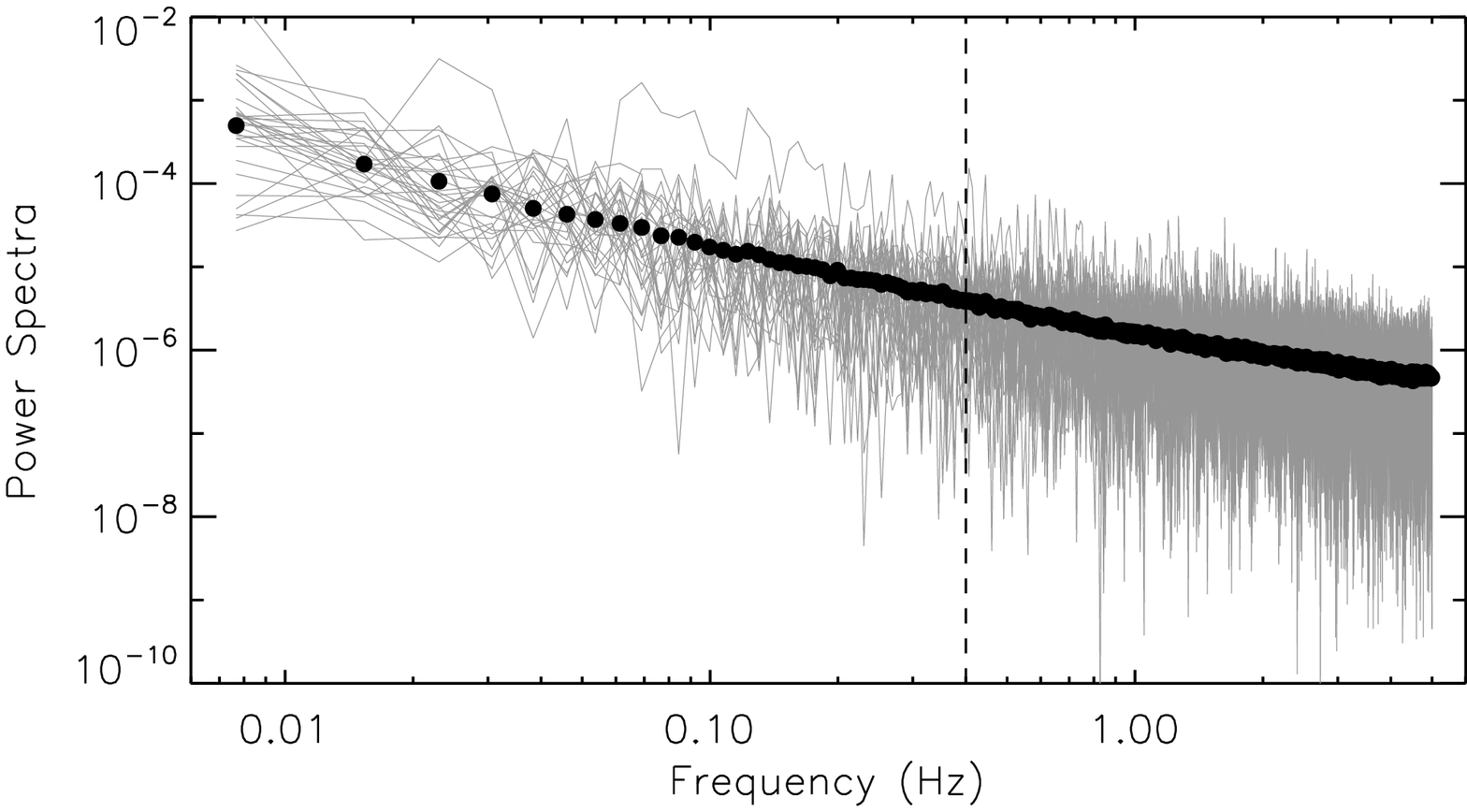}
   \end{tabular}
   \caption{The thick black dots show the median power spectrum for a
     single scan leg (4\,deg) obtained from the bulk ($32\times64$ for
     green and $16\times32$ for red) of the bolometer timelines (after
     deglitching). The top and bottom panels show the typical 100- and
     160-$\mu$m power spectra (in arbitrary units), respectively. Some
     few individual power spectra are shown with grey lines to
     visualise the scatter.  The $1/f$ signature is evident on the
     data. The length of the high-pass filters described on
     \S\,\ref{naive_proj} (3.4 and 2.5\,arcmin for green and red
     timelines, respectively) are shown in terms of frequency by
     vertical dashed lines.}
   \label{Foneoverf}
\end{figure}

Within HIPE, there are two possible ways to project the timelines of
scan-mode observations. First, a simple (also called `na\"ive' or
drizzle) projection ({\tt photProject}; see \S\,\ref{naive_proj}) on
sky for every frame -- simply dividing and weighting the signal
according to the projection of each bolometer onto a pixelated sky
(\citealt{fruchter02}, \citealt{serjeant03}). Given the pronounced
$1/f$ noise (see Fig.~\ref{Foneoverf}), an aggressive high-pass filter
has to be applied to the timelines for this task to work
efficiently. Second, the Microwave Anisotropy Dataset mapper (MADmap
-- {\tt runMadMap}) can be also used (\citealt{cantalupo10}, and see
\S\ref{SSalternative}). This task uses a maximum-likelihood map
reconstruction algorithm, which requires good knowledge of the noise
but does not require aggressive high-pass filtering. For the SDP data,
we opted to run the na\"ive projection.

\subsubsection{Na\"ive projection}
\label{naive_proj}

As already mentioned, in order to remove the thermal drifts from the
timelines, we have applied a boxcar high-pass filter ({\tt
  highpassFilter} -- HPF) on a length scale of 15 [3.4] and 25 [2.5]
frames [arcmin] to the 100- and 160-$\mu$m timelines, respectively
(note these scales correspond to $2\times{\rm D}+1$ where D is the
input value to run the task). The HPF subtracts a running median from
each readout frame, thereby removing all the large-scale structure
from the map, including thermal drifts, cosmic cirrus and extended
sources -- this na\"ive projection is only efficient to detect
point-like sources.

\begin{figure}
   \centering \includegraphics[width=8.1cm]{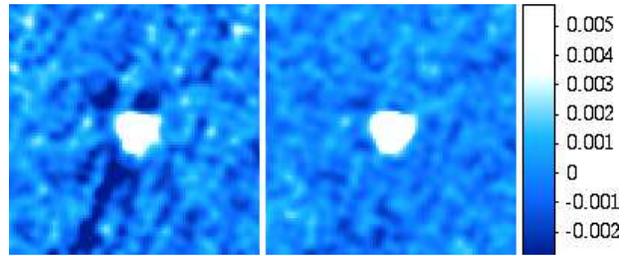} \\
   \caption{Left: the effect of aggressive high-pass filtering seen
     around a bright 100-$\mu$m source along the scan
     direction. Right: masking correction applied by the 2nd-level
     high-pass filtering approach. The colour scale is in
     Jy\,pixel$^{-1}$, where the pixel scale is 2.5\,arcsec. Note that
     the higher noise in the left image is the result of an earlier
     data reduction with an older HIPE built (for displaying purposes
     only).}
   \label{hpf_img}
\end{figure}

This aggressive high-pass filtering inevitably results in an
under-estimation of peak flux densities. Indeed, around bright sources
negative sidelobes are seen clearly along the scan-direction (see
Fig.~\ref{hpf_img}-left). For this reason, we have used the so-called
`2nd-level high-pass filtering' approach, which basically avoid a
biased median subtraction due to the presence of strong signals on the
timelines (due to real sources). We use the same mask created for the
2nd-level deglitching, i.e.\ flagging those timeline frames
contributing to all 4-$\sigma$ map pixels, in order to
not to bias the median boxcar high-pass filter estimate ({\tt
  maskname} keyword within {\tt highpassFilter}). In
Fig.~\ref{hpf_img}-right we clearly show the improvement made by this
`2nd-level high-pass filtering' which is specially important to 
co-add different scans and maintain an uniform PSF across the map.

Finally, just before producing the final images using {\tt
  photProject} we select only those frames which were used for
scanning the target GAMA field (i.e.\ removing turnarounds and
remaining calibration blocks) via the Building Block ID ({\tt BBID =
  215131301}) number.

\begin{figure*}
   \centering
   \begin{tabular}{ll}
    \includegraphics[width=17cm]{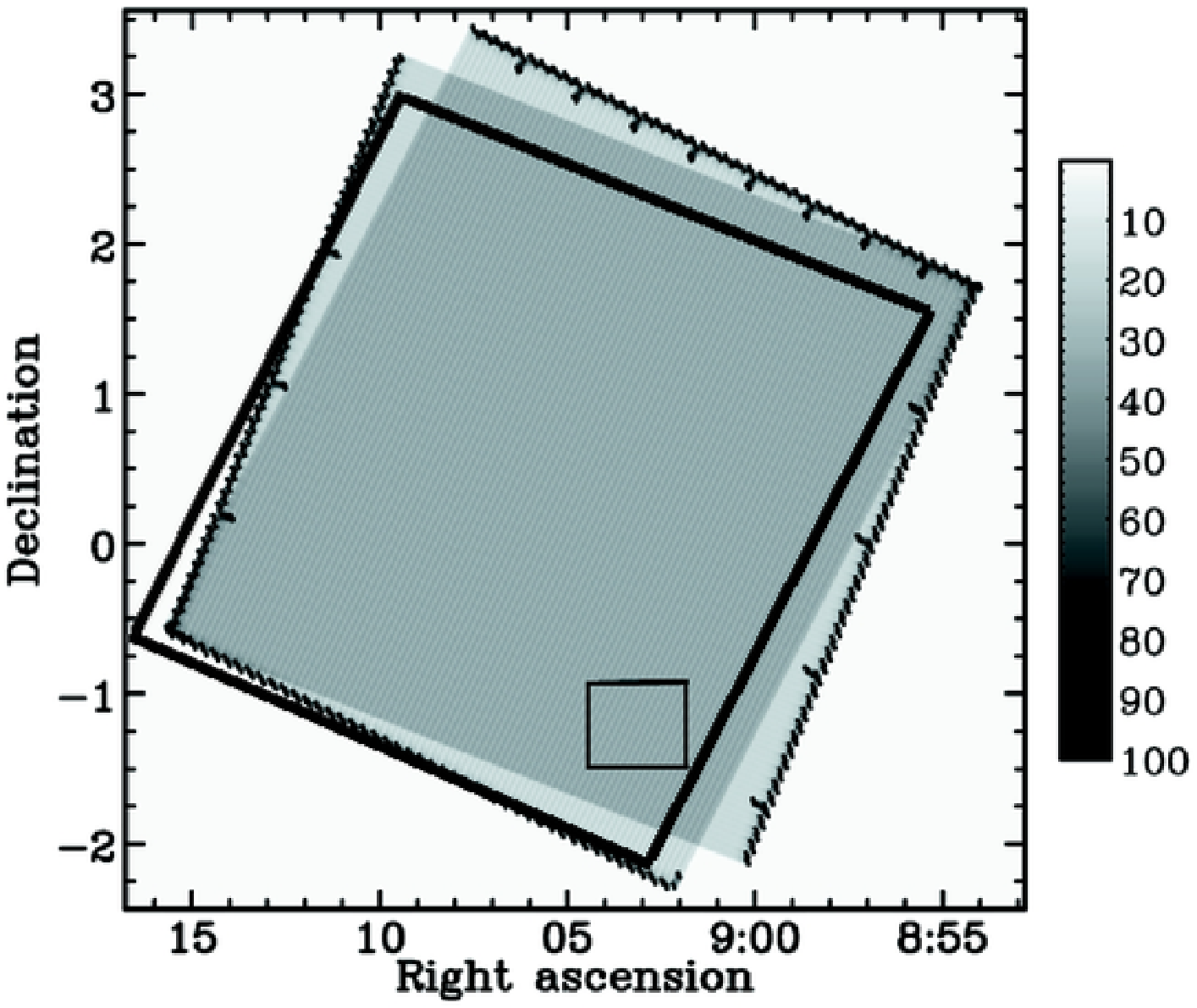}\\
    \includegraphics[width=8.5cm]{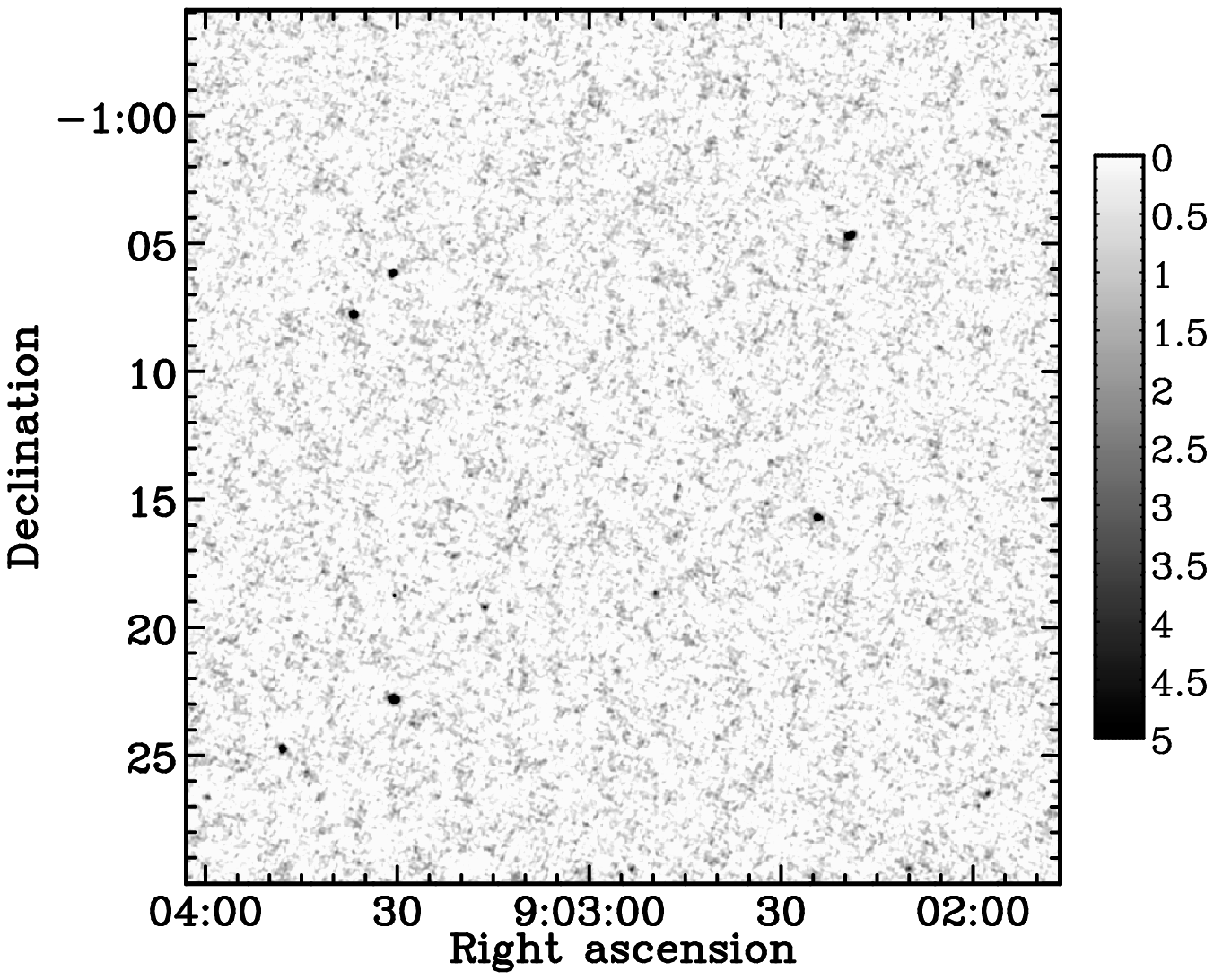} 
    \includegraphics[width=8.5cm]{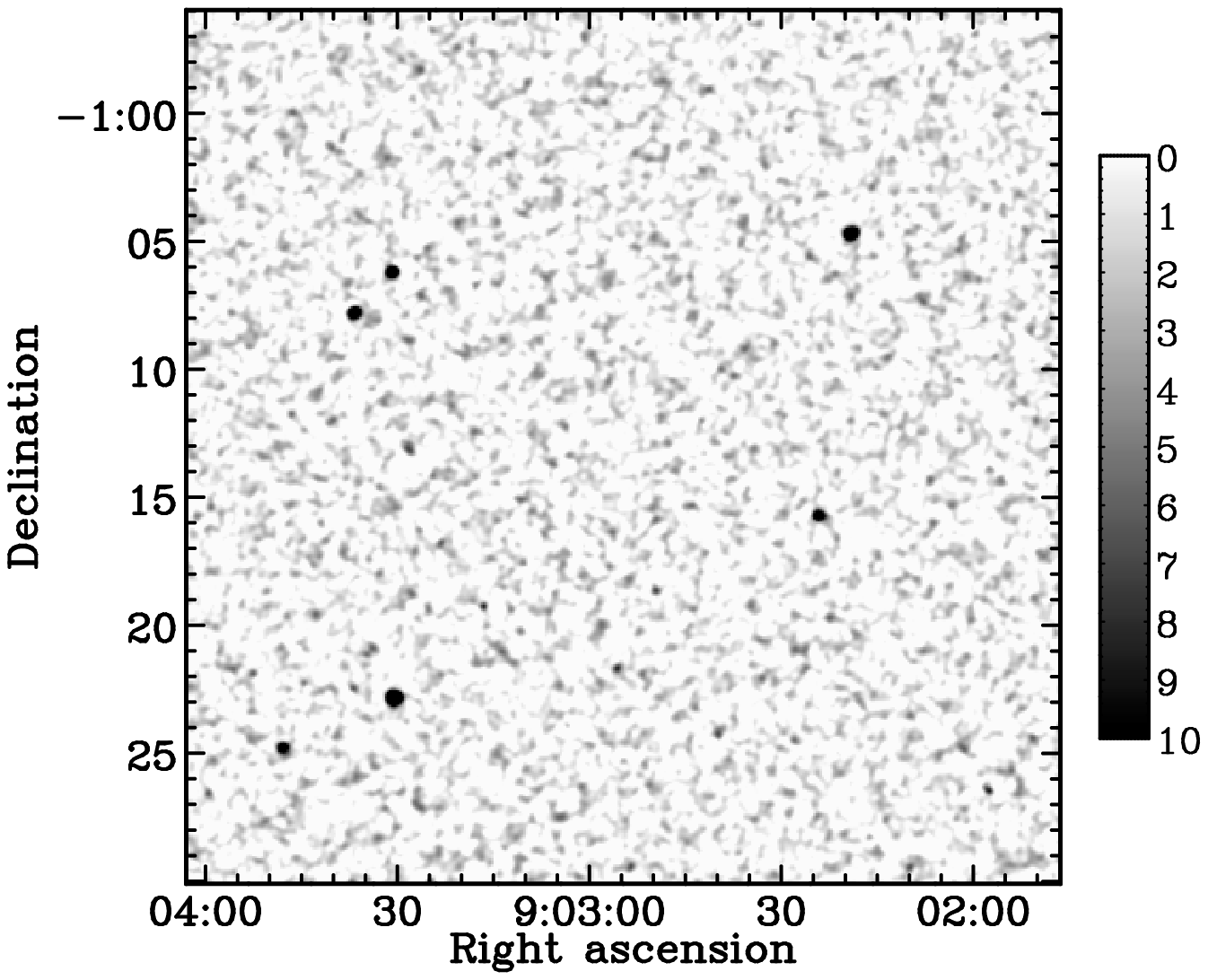} \\
   \end{tabular}
   \caption{Top: full PACS coverage for the 4$\times$4\,deg$^2$
     \hatlas\ SDP observations in units of frames per pixel (at
     160\,$\mu$m). The large, thick square shows the SPIRE coverage
     ($\sim$21\,arcmin offset). Bottom: a small $\sim0.4^2\,$deg$^2$
     region of the field (small, thin square in top figure) centred at
     R.A.\ = 9h\,2m\,54.5s and Dec.\ = $-01^\circ\,12''\,54.5'$ imaged
     at 100 (left) and 160\,$\mu$m (right). The bottom images have
     been convolved with a 2-pixel-wide Gaussian (pixels of 2.5 and
     5.0\,arcsec for green and red, respectively) for display
     purposes.}
   \label{lero_img}
\end{figure*}

The resultant maps were chosen to have 2.5- and 5.0-arcsec pixel sizes
for the green and red filters, respectively. These sizes were chosen
in consultation with the \hatlas\ SPIRE data reduction group (Pascale
et al., in prep.) and ensure that all five images can be combined
trivially (the pixel scales are 5, 10 and 10\,arcsec at 250, 350 and
500\,$\mu$m, respectively). The full PACS cross-scan coverage (top)
and a small sub-region (bottom -- a tenth of the full image) are shown
in Fig.~\ref{lero_img}.

\subsubsection{MADmap imaging}
\label{SSalternative}

Although na\"ive maps are well-suited to our early SDP science goals,
maximum-likelihood map-makers are required to recover large-scale
diffuse emission, like Galactic cirrus and/or extended local
galaxies. Here, we describe modest progress with the MADmap imaging
task ({\tt runMadMap}) implemented within HIPE.

If observations contain a good mix of spatial and temporal information
at any given point on the sky, MADmap can aid in the removal of
uncorrelated low-frequency noise (\citealt{waskett07}). We have
combined the two SDP scans (obtained $\sim$8\,hr apart;
Table~\ref{obs_table}) to produce an image that suffers less from the
pronounced drifts generated by the $1/f$ noise (see
Fig.~\ref{Foneoverf}). In order to successfully solve this inversion
method, a good characterisation of the noise must be provided to
MADmap (see \citealt{cantalupo10}). This requires that correlated
noise amongst detectors, and other correlated artifacts, must be
removed from the data, or at least mitigated.

\begin{figure}
   \centering 
   \includegraphics[width=8.1cm]{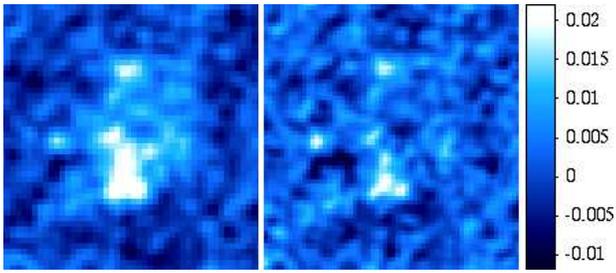} \\
   \caption{{\it Left:} an extended 160-$\mu$m source, imaged using
     MADmap ({\tt runMadMap}); {\it right:} the same source imaged
     using {\tt photProject}. Colour scale in Jy\,pixel$^{-1}$ (though
     we have not fully tested calibration of the MADmap images). The
     recovery of extended emission in the MADmap image is evident.}
   \label{Fmadmap}
\end{figure}

Fig.~\ref{Fmadmap} shows how MADmap improves the recovery of extended
emission, avoiding the loss of signal which results from the high-pass
filter required for na\"ive projection. The images show a bright,
extended source (J090402.9+005436; Thompson et al., in
preparation). The MADmap images have a smoother background than those
produced using {\tt photProject}. However, we find that MADmap
projection is sensitive to the sudden jumps produced by
multi-pixel/multi-frame glitches, and by the long thermal drifts
observed after calibration blocks (correlated features). In the future
we will present a more detailed pre-processing analysis required to
use MADmap within HIPE, and explore other approaches for imaging.

\section{Image analyses}
\label{Sanalyses}

\subsection{Point-spread function}

The observed PSF in fast-scan parallel mode (60\,arcsec\,s$^{-1}$)
suffers from strong smearing effects due to the averaged sampling
frequency and the detector time constant which results in an
elongation of the PSF in the direction of the scan. The expected {\sc
  fwhm} broadening factors are $\sim$1.9 and $\sim$1.4 with respect to
that observed using nominal scan speed (20\,arcsec\,s$^{-1}$) at 100
and 160\,$\mu$m, respectively (see ICC report in footnote~\ref{fnlutz}
for more details). The PSF shape becomes even more complicated when
combining different scan directions. We have roughly modelled the PSF
based on observations to the Vesta (OD160) asteroid (provided by
ICC). We take this image and co-add it to the same image but rotated
by 90$^\circ$ to simulate the cross-scanning. A 2-D Gaussian fit
resulted in a {\sc fwhm} of 8.7 and 13.1\,arcsec at 100 and
160\,$\mu$m, respectively.

In an attempt to measure the PSF using the bright sources in our final
image product, we have stacked the PACS signal of 25 radio sources
detected in the Faint Images of the Radio Sky at Twenty Centimeters
\citep[FIRST --][]{becker95} survey. The measured {\sc fwhm} of the
stacked signal using a 2-D Gaussian fit results in $11.25\times12.25$
and $15\times17.5$\,arcsec$^2$ at 100 and 160\,$\mu$m,
respectively. These are larger than the PSF produced from the Vesta
images, which could be the result of slight pointing offsets, time
shifts on science data, intrinsically extended and/or blended
sources. Given these reasons and the uncertainty on the observed PSF,
we have considered the fits provided by the PACS ICC (see
footnote~\ref{fnlutz}) to be inappropriate for our specific
\hatlas\ analyses.

\begin{figure}
   \centering
    \includegraphics[width=7.8cm]{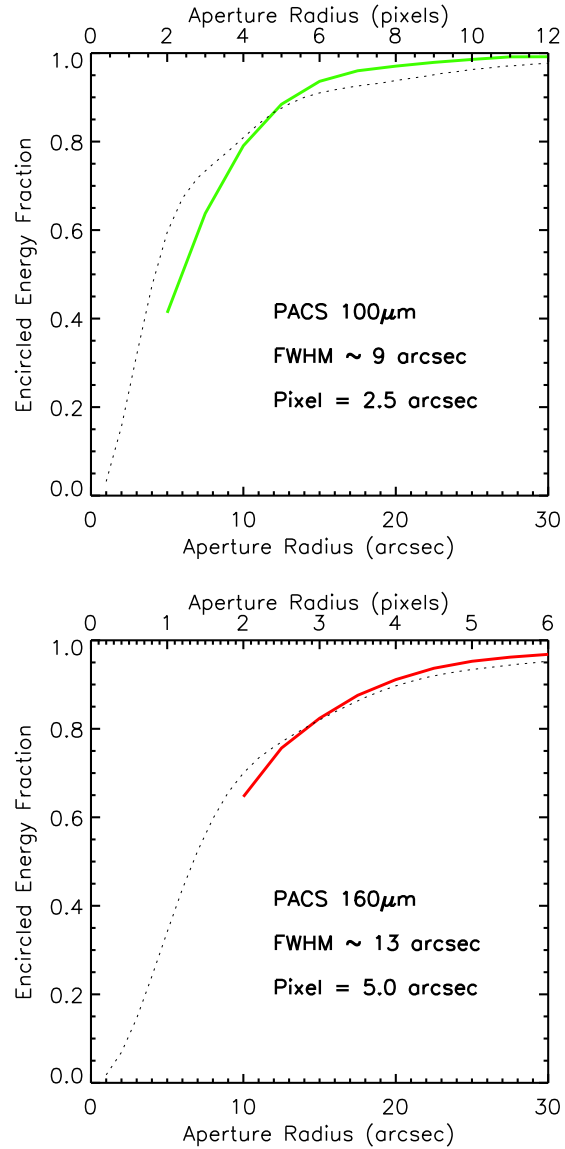}
    \caption{Encircled energy fraction normalised to an aperture of
      60\,arcsec (effectively zero background subtraction) as a
      function of aperture radius, in steps of 2.5\,arcsec, for both
      PACS 100- (left) and 160-$\mu$m (right) passbands. The estimates
      are based on a bright point-like source found in an ICC
      calibration observations (OBSID 1342190267 and 1342190268) of
      $\alpha$~Bootes made in fast-scan SPIRE/PACS parallel mode, with
      two cross-scans. Dotted lines correspond to the ICC estimates,
      derived from slow-scan observations (OD160) of Vesta (normalised
      to a same 60\,arcsec aperture). The quoted pixel-scales are the
      ones used for the map production.}
    \label{Fap_corr}
\end{figure}

We have instead adopted an empirical approach to characterise the
PSF. We use a bright, point-like source (flux densities of $S_{\rm
  100\mu m}= $\ 7.5\,Jy and $S_{\rm 160\mu m}= $\ 2.9\,Jy) detected
near the field centre of a calibration observation (OBSID 1342190267
and 1342190268) made in the $\alpha$~Bootes field. These data were
observed in the same fast-scan parallel mode as our \hatlas\ data
(with similar cross-scans) and reduced using exactly the same pipeline
described in this paper. Using this bright source we are able to
describe the radial profile of the PSF (and necessary aperture
corrections) empirically, as shown in Fig.~\ref{Fap_corr}. We follow
the same procedure as the PACS ICC, normalising our measures to a
radius aperture of 60\,arcsec, with background subtraction done in an
annulus between radius 61 and 70\,arcsec (effectively zero in our
map). We find a good agreement between our 160-$\mu$m profile and that
taken in slow-scan mode. As expected, we find evidence that small
apertures under-estimate the encircled energy in fast-scan mode
(especially for the green filter). With larger apertures we obtain
smaller aperture corrections, indicating that our approach
under-estimates the size of the PSF wings -- standard empirical
measurements could never detect the broad wings of the
PSF. Fig.~\ref{Fap_corr} is used for \hatlas\ source extraction, as
described by Rigby et al.\ (in prep.).

\subsection{Sensitivity of the maps}
\label{SSsensitivity}

Several different calibration files have been implemented in HIPE. We
use the default `version 3' of the flux calibration files from HIPE
v3.0.859 which have been found to be biased by the PACS
ICC\footnote{\label{fnscanning} PICC-ME-TN-035, February 23, 2010,
  version 1.1 report by The PACS ICC.}. Flux densities have been found
to be over-estimated by a factor of 1.09 (at 100\,$\mu$m) and 1.29 (at
160\,$\mu$m) with respect to previous observations by {\it Spitzer}
and the {\it Infrared Astronomical Satellite (IRAS)}. These
corrections are applied on the public release of these maps. The
absolute calibration uncertainties measured by the ICC calibration
campaign are currently within 10 and 20 per cent for the 100- and the
160-$\mu$m wavebands, respectively.

\begin{figure}
   \centering \includegraphics[width=8.1cm]{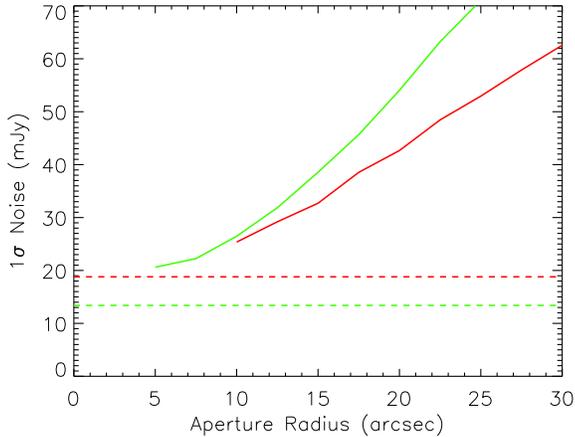}
    \caption{Continuum lines: noise estimates for the 100- (green) and
      160-$\mu$m (red) images as a function of aperture size. Dashed
      lines: the 1-$\sigma$ point-source sensitivities expected for
      \hatlas\ based on the HSpot observation planning tool (as shown
      in \citealt{eales10}).}
    \label{Fnoise_corr}
\end{figure}

To measure the noise in the maps we have used the aperture corrections
from Fig.~\ref{Fap_corr} in combination with the calibration
correction factors stated above. We made aperture measurements at
random locations within the central part of the image -- placing
apertures randomly and measuring counts within those apertures. To
ensure that our noise measurements were not affected by sources, we
made 10,000 aperture measurements, then carried out iterative clipping
at the 2.5-$\sigma$ level, where we take the median deviation
measurements and throw out points more than $\pm$2.5$\sigma$ from the
median. We repeat this process until it converges -- in typically 3--7
iterations. As a cross-check, we also fit a Gaussian to the clipped
histogram of the data distribution (e.g.\ from $-2.5$- to 0.5- or
1-$\sigma$, re-deriving $\sigma$ and repeating) which is better for
confused maps. In this case, simple $\sigma$ clipping and fitting the
histogram distribution yield the same noise level. Given that the PACS
data are filtered, the local background is fairly flat and we do not
need to subtract a background torus for each of the
measurements. These estimates are shown in Fig.~\ref{Fnoise_corr}.

We show the noise is strongly dependent on the aperture we use.
Fig.~\ref{Fnoise_corr} may suggest more correlated noise in the
100-$\mu$m data given the larger increase in noise as a function of
spatial scale compared with the 160-$\mu$m data. We find that the
1-$\sigma$ noise varies from 25 to 33\,mJy for 9 to 13\,arcsec, and
from 30 to 48\,mJy for 13 to 18\,arcsec aperture radii for the 100-
and 160-$\mu$m wavebands, respectively. These measurements are less
sensitive than those predicted by HSpot (13.4 and 18.8 for green and
red, respectively) and quoted in \citet{eales10}. The sensitivities
predicted by HSpot are based on the assumption that the noise power
spectrum at 3\,Hz should be white. In Fig.~\ref{Foneoverf} we can
clearly see that when we use a high-pass filter to tackle timeline
drifts, the remaining noise at 3\,Hz is not white: it includes other
noise components that must reduce the sensitivity of the final
images. A further analysis using a different imaging approach may be
required.

\subsection{Flux density calibration}

We performed a sanity check on the flux calibration making use of
photometric 100-$\mu$m IRAS coverage in the field. We select 14 IRAS
sources (robustly detected at 100\,$\mu$m) with clear detections in
our PACS image. A full description of the source extraction for the
\hatlas\ survey is presented in Rigby et al.\ (in preparation). For
this test, we perform a source extraction using aperture photometry
within SEXtractor ({\tt flux auto}) and correcting flux densities
using Fig.~\ref{Fap_corr}. We find a bootstrapped median of $(S_{\rm
  100-PACS}/S_{\rm 100-IRAS}) = 1.03\pm0.08$ for the flux density
ratio. Despite the small number of sources, this roughly confirms the
quality of the flux calibration within HIPE. According to an ICC
report (footnote~\ref{fnscanning}), these calibrations are still under
development within HIPE.

Note that monochromatic flux densities quoted from broadband
photometry are dependent on the shape of the SED (colour
corrections). Indeed, based on the PACS filter profiles (see
Fig.~\ref{filters}), small variations of the order of $\ls5$\,per cent
have to be applied to the observed flux density at the reference
wavelength. These variations are significant for cold ($<$20\,{\sc k})
sources (\citealt{poglitsch10}).

\subsection{Astrometry}

The pipeline-reduced maps are already astrometrically calibrated. We
confirmed and checked the accuracy of the astrometric solution of the
PACS green and red maps against the FIRST catalogue, the parallel
SPIRE catalogue (Rigby et al., in preparation) and with respect to
each other. Catalogue sources were deemed to be matches if an
association was found within 6\,arcsec. The mean offsets are
summarised in Table~\ref{tab:ast} along with the number of matches
between the input catalogues used to determine the offsets. The mean
offsets and standard deviations are smaller than the search radius
used for the matching. This, together with the negligible sub-pixel
change in astrometric offset for association radii between 5 and
10\,arcsec, supports the idea that the quoted offsets and standard
deviations are representative of the astrometric accuracy of the
maps. Both green and red maps were seen to be well aligned with each
other and to the SPIRE and FIRST catalogues. An analysis of the
direction of the offsets suggests that there is a systematic offset
between the PACS green map and SPIRE catalogue of $-$1 and
+2.4\,arcsec in R.A.\ and Dec., respectively.  This might be related
to recent ICC findings relating to a 50-ms time shift in science data
(T.\ M\"uller, priv.\ comm.). As all of the mean offsets are close to
the size of a single PACS pixel (and significantly smaller than an
individual SPIRE pixel -- see \S\,\ref{Sprojection}) we have not
applied any global offsets to the maps but simply quote the offsets
and accuracy between frames here.

\begin{table}
   \centering
   \begin{tabular}{@{} lccccc @{}} 
\hline
\hline
 & R.A. & r.m.s.\ & Dec. & r.m.s.\ & $N$\\
 & offset & (arcsec) & offset & (arcsec) & \\
 & (arcsec) &  & (arcsec) &  & \\
 \hline
  100$\mu$m --   160$\mu$m & -0.5$\pm$0.1 & 1.2 &  1.3$\pm$0.1 & 1.4 & 94\\
  100$\mu$m -- SPIRE    & -1.0$\pm$0.1 & 1.4 &  2.4$\pm$0.1 & 1.3 & 93 \\
  100$\mu$m -- FIRST    & -1.3$\pm$0.3 & 1.5 &  2.1$\pm$0.4 & 0.8 & 25\\
\hline
  160$\mu$m --   100$\mu$m &  0.5$\pm$0.1 & 1.2 & -1.3$\pm$0.1 & 1.4 & 94\\
  160$\mu$m -- SPIRE    & -0.6$\pm$0.2 & 1.8 &  1.1$\pm$0.1 & 1.5 & 138\\
  160$\mu$m -- FIRST    & -0.8$\pm$0.3 & 1.8 &  0.8$\pm$0.2 & 1.1 & 29\\
\hline
SPIRE - FIRST        &  0.1$\pm$0.2 & 2.0 &  0.1$\pm$0.2 & 2.0 & 85 \\
\hline
\hline
   \end{tabular}
   \caption{A table summarising the mean astrometric offsets and
     r.m.s\ deviations between the PACS images and other
     catalogues. $N$ stands for the number of matched sources. The
     SPIRE-FIRST comparison is included for completeness.}
   \label{tab:ast}
\end{table}

\section{Concluding remarks}
\label{Sconclusions}

We have been able to produce science-quality images of a large
(4$\times$4\,deg$^2$) area of sky with the PACS instrument on board
the {\it Herschel Space Observatory}. We describe the data processing
implemented within HIPE and used to image the SDP region of the
\hatlas\ survey. Data were taken in fast-scan (60\,arcsec\,s$^{-1}$)
parallel mode using PACS at 100 and 160\,$\mu$m.

During the data reduction we faced many difficulties, principally due
to the large volume of data. Only machines with at least 60\,Gbytes of
RAM were able to process the full data reduction. The pipeline was
developed using customised HIPE procedures from data retrieval to
final imaging (see Fig.~\ref{Fpipeline}). We describe an effective
approach to tackle powerful glitches and the pronounced $1/f$ noise
present in the data. We perform a careful analysis to protect the
signal at the position of bright sources using masks during
deglitching and filtering processes. In this early SDP data reduction,
we cannot guarantee the complete absence of glitches in the image
products. This could result in spurious sources, with obvious
consequences when, e.g., determining accurate source counts or
searching for outliers in flux or colours. The following \hatlas\ data
releases will mitigate this problem.

Based on the final images, we describe the PSF, sensitivity,
calibration and astrometric quality of the maps. In particular, we
find previous HSpot sensitivities are too optimistic (see
Fig.~\ref{Fnoise_corr}) compared to those measured in our maps. In
principle, the origin of this discrepancy is unknown and further
development of the pipeline may be required to allow us to reach a
sensitivity closer to the instrumental noise expectations.

\hatlas\ PACS SDP products are avaiable at {\tt
  http://www.h-atlas.org/}.

\section*{Acknowledgements}

PACS has been developed by a consortium of institutes led by MPE
(Germany) and including UVIE (Austria); KU Leuven, CSL, IMEC
(Belgium); CEA, LAM (France); MPIA (Germany); INAFIFSI /OAA/OAP/OAT,
LENS, SISSA (Italy); IAC (Spain). This development has been supported
by the funding agencies BMVIT (Austria), ESA-PRODEX (Belgium),
CEA/CNES (France), DLR (Germany), ASI/INAF (Italy), and CICYT/MCYT
(Spain). We would like to thank the PACS-ICC team for providing
excellent support to the \hatlas\ project and for the various HIPE
developments that comprise the current pipeline. Finally, we thank the
referee for comments that significantly improved this paper.

\setlength{\labelwidth}{0pt} 

\bibliographystyle{mn2e}
\bibliography{refs}

\end{document}